\documentclass[prb,preprint, amsmath,amssymb]{revtex4-1}

\usepackage{graphicx}
\usepackage[section]{placeins}
\usepackage{soul}

\usepackage{hyperref}   
\hypersetup{
	colorlinks=true, 
	linkcolor=blue, 
	citecolor=blue, 
	urlcolor=magenta, 
	pdfborder={0 0 0} 
}

\begin{document}
\title{Geometric invariance from outer surfaces: Laplace-governed magnetization in the high-permeability limit}

\author{Yujun Shi}
\email{yujunshi@sxu.edu.cn}
\affiliation{Collage of Physics and Electronic Engineering, Shanxi University, Taiyuan 030006, China}

\begin{abstract}
	The magnetization of bodies in static fields is a textbook topic in electrodynamics, governed by Laplace equations with interface continuity (transmission) conditions. In the infinite-permeability limit, textbooks emphasize the quasi-equipotential interior and normality of the external field at the boundary, but leave the exterior largely uncharacterized. Here we identify a singular property that has not been explicitly stated in the existing literature: in this limit, the entire external magnetic response, including the external field distribution and all multipole moments, is determined solely by the outer surface geometry, independent of internal structure or deformation. Numerical simulations confirm this limiting property is well approximated under finite high-permeability conditions, thereby providing a theoretical basis for the lightweight design of magnetic devices such as flux concentrators. Since analogous Laplace transmission problems arise across physics, including heat conduction, electrostatic polarization, and acoustic scattering, this geometric invariance exhibits cross-disciplinary universality. Together with the quasi-equipotential property, it provides a complementary and essentially complete characterization of Laplace transmission problems in the infinite-permeability limit.
	
\end{abstract}

\maketitle

\section {Introduction}\label{sec1}

Magnetization in static fields reduces to solving Laplace equations with interface continuity (transmission) conditions~\cite{griffiths2014introduction,jackson_classical_1999,Zangwill_2012}, a class of problems often referred to in the partial differential equations (PDE) literature as Laplace transmission problems. In the limit of infinite permeability ($\mu \to \infty$), the body is nearly equipotential, and textbooks usually highlight that the external field is normal to the surface~\cite{griffiths2014introduction,jackson_classical_1999,Zangwill_2012,cheng1989field}. Yet this raises a fundamental question: does the exterior field admit a deeper invariant structure?

In physics and mathematics, transformations and invariances (symmetries) form the backbone of our understanding. They not only guide the formulation of natural laws and mathematical classification, but also serve as conceptual anchors that organize our reasoning. In the high-permeability limit, a body becomes quasi-equipotential, displaying an invariance that persists under arbitrary geometric deformations. Given that the interior and exterior potential fields are tightly coupled via boundary conditions on the outer surface, the exterior potential field may likewise exhibit a form of geometric invariance.

We show that this invariance indeed exists. In the $\mu \to \infty$ limit, the exterior potential field is an outer-surface determined invariant: it depends solely on the body's outer boundary, regardless of internal geometry. Consequently, the field distribution in the exterior region and all multipole moments of the body depend solely on the body’s outer boundary. Notably, in electrostatics, an analogous statement about perfect conductors is so obvious that it is often left unstated. In magnetostatics, however, this invariance cannot be directly inferred from the case of perfect conductors. For perfect conductors, the boundary conditions are not obtained from the general interfacial continuity relations, but instead from an additional constraint that the electric field vanishes inside the conductor. In fact, the truly symmetric counterpart of the magnetostatic problem with infinite permeability is the electrostatic problem involving materials with infinite dielectric permittivity. This point will be discussed in more detail in Sec.~\ref{sec2}.

At first sight, this invariance appears to belong to the level of well-established textbook knowledge. However, we have not found it explicitly stated in any of the standard references~\cite{griffiths2014introduction,jackson_classical_1999,Zangwill_2012,cheng1989field,vanderlinde2004classical,Purcell_Morin_2013,sommerfeld1952electrodynamics,portis1978electromagnetic,10.1119/1.1937585}, nor has it been recognized or utilized as a general design principle in practical applications. 
As discussed in Sec.~\ref{Application in MFC}, for instance, magnetic flux concentrators intended to enhance local magnetic fields are conventionally constructed as solid bodies, 
whereas our analysis indicates that a lightweight hollow structure would perform equivalently in the high-permeability limit. In 2014, Ciric establishes a single-valued scalar-potential formulation for magnetostatic problems involving ideal ferromagnetic bodies ($\mu \to \infty$). However, he does not address the geometric invariance of the exterior field or the total magnetic moment with respect to the bodies’ interior geometry\cite{6587817}. This observation points to a rather surprising conclusion: for more than a century since the classical framework of electromagnetism was established, magnetization problems have been regarded as well understood. Nevertheless, the high-permeability limit ($\mu \to \infty$) and its associated geometric invariance---namely, that the external magnetic response is determined solely by the outer boundary---has received surprisingly little explicit attention. Numerical simulations demonstrate that this limiting property is well approximated under finite high-permeability conditions, confirming its robustness and practical relevance.

\begin{figure}[h!]
	\includegraphics[width = 1\textwidth]{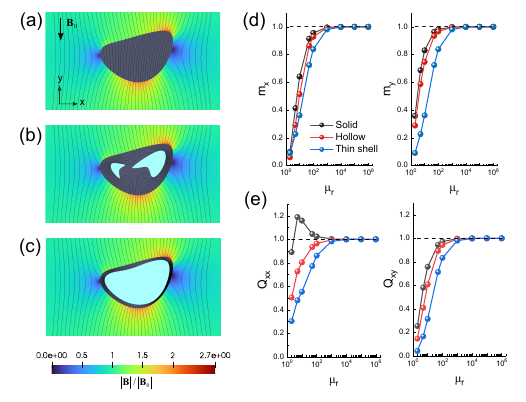}
	\caption{External magnetic response of bodies with identical outer surfaces under a uniform magnetic field.  
		(a--c) Magnetic flux lines and their density in the exterior of three bodies with identical outer surfaces but differing internal structures, each with a relative permeability of $10^6$ : a solid body (a), a hollow shell (b), and a thin shell (c).  
		(d--e) Magnetic dipole vector components (d) and quadrupole tensor components (e) as functions of increasing permeability. All values are normalized to those of the solid body at relative permeability $10^6$.}
	\label{Figure02}
\end{figure}

The paper is organized as follows. In Sec.~\ref{sec2}, we present the formulation of the magnetization problem for linear media in arbitrary static magnetic fields. We compare the magnetization problem of materials with infinite permeability to the electrostatic problem of a conductor, and provide the core proof of the geometric invariance reported in this work (the complete derivation is given in the Appendix for clarity). In Sec.~\ref{sec3}, we perform a series of two-dimensional finite-element simulations to demonstrate the geometric invariance. In Sec.~\ref{Application in MFC}, using the hollow design of a magnetic flux concentrator as an example, we illustrate how this invariance can be applied to achieve lightweight device designs. In Sec.~\ref{sec5}, we discuss the pedagogical value of this invariance and its possible implementation in teaching. Readers who are not particularly interested in the rigorous mathematical proof may skip the derivation and focus on the main conclusion and numerical demonstrations of the invariance.

\section{Problem Formulation and Main Results}\label{sec2}

\subsection{Magnetization problem in arbitrary static magnetic fields}

Consider a bounded region \( \Omega \subset \mathbb{R}^d \) (\( d = 2,3 \)) representing a homogeneous, isotropic magnetic body with a linear constitutive relation \( \mathbf{B} = \mu \mathbf{H} \), embedded in an unbounded medium with permeability \( \mu_0 \).
Here, \( \Omega \) denotes the magnetic body, while the surrounding two- or three-dimensional space extends to infinity. Without free currents, the magnetic field derives from a scalar potential \( \phi \), such that \( \mathbf{H} = -\nabla \phi \). The divergence-free condition \( \nabla \cdot \mathbf{B} = 0 \) leads to a classical transmission problem for the Laplace equation:
\begin{equation}\label{transmission problem}
	\Delta \phi^- = 0 \quad \text{in } \Omega, \qquad 
	\Delta \phi^+ = 0 \quad \text{in } \mathbb{R}^d \setminus \overline{\Omega},
\end{equation}
where the superscripts ``$-$'' and ``$+$'' denote the interior and exterior of the body, respectively,  
and $\mathbb{R}^d \setminus \overline{\Omega}$ denotes the exterior domain excluding the boundary $\partial\Omega$.

The continuity (transmission) conditions on the boundary \( \Gamma = \partial \Omega \) are
\begin{align}
	\mathbf{n} \times (\mathbf{H}_\text{out} - \mathbf{H}_\text{in}) &= 0, \\
	\mathbf{n} \cdot (\mathbf{B}_\text{out} - \mathbf{B}_\text{in}) &= 0,
\end{align}
which are equivalent, in terms of the scalar magnetic potential, to
\begin{equation}\label{interface conditions}
	\phi^- = \phi^+ \quad \text{on } \Gamma, \qquad
	\mu \frac{\partial \phi^-}{\partial \mathbf{n}} = \mu_0 \frac{\partial \phi^+}{\partial \mathbf{n}} \quad \text{on } \Gamma.
\end{equation}
To ensure a well-posed exterior problem, the potential is required to satisfy the asymptotic condition
\begin{equation}\label{asymptotic condition}
	\phi^+(\mathbf{r}) - \phi^{\mathrm{bg}}(\mathbf{r}) \to 0 
	\quad \text{as } |\mathbf{r}| \to \infty ,
\end{equation}
where \( \phi^{\mathrm{bg}} \) denotes a prescribed harmonic background potential representing the applied magnetic field. This condition guarantees that the perturbation induced by the object vanishes far away from it.

\subsection{Comparison with the electrostatic problem of a conductor}

It is well known that the magnetization problem for materials with infinite magnetic permeability is closely analogous to the electrostatic behavior of a perfect conductor. Both are governed by  the Laplace equation, though subject to different boundary conditions. In electrostatic equilibrium, the electric field vanishes inside the conductor, rendering it an equipotential body. Instead of the continuity conditions given in Eqs.~(\ref{interface conditions}), the boundary of a conductor enforces a strict equipotential constraint:
\begin{equation}\label{equipotential condition}
	\phi^- = \phi^+ \quad \text{on } \Gamma, \qquad
	\phi^-=\text{const.}   \quad \text{in }\Omega,
\end{equation}
which is equivalent to the following interface conditions:
\begin{equation}\label{limit interface condition for conductor}
	\phi^- = \phi^+ \quad \text{on } \Gamma, \qquad
	\frac{\partial \phi^-}{\partial \mathbf{n}} = 0 \quad \text{on } \Gamma.
\end{equation}

While in the magnetization problem of a material with infinite magnetic permeability, the continuity conditions given by Eqs.~(\ref{interface conditions}) reduce to:
\begin{equation}\label{limit interface condition for infinite permeability}
	\phi^- = \phi^+ \quad \text{on } \Gamma, \qquad
	\frac{\partial \phi^-}{\partial \mathbf{n}} = \frac{\mu_0}{\mu} \frac{\partial \phi^+}{\partial \mathbf{n}}\to 0 \quad \text{on } \Gamma.
\end{equation}

From the mathematical structure alone, it is evident that the electrostatic problem of a perfect conductor can be viewed as the limiting case of the magnetostatic problem for a material with infinite magnetic permeability. However, caution must be taken: this limiting correspondence implies analogy, not full mathematical equivalence or symmetry. In fact, the truly symmetric counterpart of the magnetostatic problem with infinite permeability is the electrostatic problem involving materials with infinite dielectric permittivity.

For perfect conductors, the boundary conditions are not derived from the general interface conditions for the electric field, analogous to Eqs.~(\ref{interface conditions}), but rather follow from an additional constraint that the electric field inside the conductor is zero. In contrast, in electrostatic problems involving materials with infinite dielectric permittivity, or in magnetostatic problems with infinite magnetic permeability, the internal electric or magnetic field approaches zero only asymptotically and is not strictly zero. This distinction is fundamental, setting these cases apart from the electrostatics of perfect conductors.

To further illustrate this distinction, consider the total magnetic moment of the body:
\begin{equation}
	\mathbf{m} = \int_{\Omega} \mathbf{M}~dV = \left( \frac{\mu}{\mu_0} - 1 \right) \int_{\Omega} \mathbf{H}~dV.
\end{equation}
Evidently, if the internal magnetic field vanishes exactly, the total magnetic moment must also vanish. Alternatively, the magnetic moment can also be expressed in the following equivalent form~\cite[Jackson, page~3, \emph{``Theorems from Vector Calculus''}]{jackson_classical_1999}:
\begin{equation}\label{M_potential}
	\begin{aligned}
		\mathbf{m}
		&= \left( \frac{\mu}{\mu_0} - 1 \right) \int_{\Omega} \mathbf{H}~dV \\
		&= -\left( \frac{\mu}{\mu_0} - 1 \right) \int_{\Omega} \nabla \phi~dV \\
		&= -\left( \frac{\mu}{\mu_0} - 1 \right) \int_{\partial\Omega} \phi~\mathbf{n}~dS,
	\end{aligned}
\end{equation}
where \(\mathbf{n}\) is the outward unit normal vector on the surface \(\partial \Omega\). This further implies that if the boundary is held at a strict equipotential, the surface integral vanishes. This follows from the identity that the surface integral of the outward unit normal vector over any closed surface is zero, i.e.,
\begin{equation}
\int_{\partial \Omega} \mathbf{n}~dS = \mathbf{0},
\end{equation}
which in turn leads to a vanishing total magnetic moment. Therefore, although the magnetostatic behavior of materials with infinite permeability is often compared to that of perfect conductors in electrostatics, the analogy is not exact. 

The geometric invariance determined by the outer surface, as identified in this work, is evidently valid for the electrostatic problem of perfect conductors. Unlike the purely outer-surface-confined nature of electrostatic induction in conductors, the magnetic response depends on the volumetric distribution of magnetization, which is associated with a vanishingly small yet nonzero internal magnetic field, and is therefore sensitive to the internal geometry of the body. This is more explicitly reflected in Eqs.~(\ref{M_potential}):
\[
\begin{aligned}
	\mathbf{m}
	&= -\left( \frac{\mu}{\mu_0} - 1 \right) \int_{\partial\Omega} \phi\, \mathbf{n}\, dS\\
	&= -\left( \frac{\mu}{\mu_0} - 1 \right) \left( \int_{\partial\Omega^{\text{out}}} \phi\, \mathbf{n}\, dS + \int_{\partial\Omega^{\text{in}}} \phi\, \mathbf{n}\, dS \right)
\end{aligned}
\]
showing that the total magnetic moment includes contributions from both the outer and inner surfaces. The inner-surface integral is generally nonzero. Thus, even in the infinite-permeability limit, the independence of the total magnetic moment from the internal geometry is not self-evident, and not obvious.

\subsection{Core argument of the proof}

The outer-surface determined invariance can be understood most transparently within a boundary-integral formulation based on Green's functions. The complete derivation is presented in the Appendix section for clarity. The core of the proof can be summarized as follows:  As $\mu/\mu_0 \to \infty$, the interior becomes quasi-equipotential. Since only potential differences are physically meaningful, we can choose the reference level such that the interior potential tends to zero, yielding $\varphi^+|_{\partial\Omega^{out}} \to 0$. The exterior thus reduces to a Dirichlet exterior problem governed solely by quasi-zero potential prescribed on the outer surface:

\begin{equation}\label{Dirichlet exterior problem}
	\begin{cases}
		\Delta\phi^+(\mathbf{r})=0,&\mathbf{r}\in\mathbb{R}^d\setminus\overline{\Omega}\\ \phi^+(\mathbf{r})\to0,&\mathbf{r}\in\partial\Omega^{out}\\\phi^+(\mathbf{r})\to\phi^{bg}(\mathbf{r}),&\left|\mathbf{r}\right|\to\infty
	\end{cases}
\end{equation}
The formal solution of Eqs.~(\ref{Dirichlet exterior problem}) can be expressed in terms of a boundary integral involving the Green's function\cite[Jackson, sec.~1.10, \emph{``Formal Solution of Electrostatic Boundary-Value Problem with Green Function''}]{jackson_classical_1999},
\begin{equation}\label{external formal solution of potential}
	\begin{aligned}
		\phi^+(\mathbf{r})&=-\int_{\partial\Omega^{out}}dS^{\prime}u(\mathbf{r}^{\prime})\frac{\partial G(\mathbf{r}-\mathbf{r}^{\prime})}{\partial\mathbf{n}^{\prime}}+\phi^{bg}(\mathbf{r})\\
		&\to\int_{\partial\Omega^{out}}dS^{\prime}\phi^{bg}(\mathbf{r}^{\prime})\frac{\partial G(\mathbf{r}-\mathbf{r}^{\prime})}{\partial\mathbf{n}^{\prime}}+\phi^{bg}(\mathbf{r})
	\end{aligned}
\end{equation}
where \(u(\mathbf{r})\) is the perturbation term, defined as 
\[\phi^+(\mathbf{r})=u(\mathbf{r})+\phi^{bg}(\mathbf{r}),
\]
and \(G(\mathbf{r}-\mathbf{r}^{\prime})\) is the existing and unique Green's function satisfying
\begin{equation}\label{external Green's function}
	\begin{cases}
		\Delta G(\mathbf{r}-\mathbf{r}^{\prime})=-\delta(\mathbf{r}-\mathbf{r}^{\prime}),\quad\mathbf{r},\mathbf{r}^{\prime}\in\mathbb{R}^d\setminus\overline{\Omega}\\
		G(\mathbf{r}-\mathbf{r}^{\prime})=0,\quad\mathbf{r}\in\partial\Omega^{out},\mathbf{r}^{\prime}\in\mathbb{R}^d\setminus\overline{\Omega}\\
		G(\mathbf{r}-\mathbf{r}^{\prime})\to0,\quad|\mathbf{r}-\mathbf{r}^{\prime}|\to\infty,\mathbf{r}\in\partial\Omega^{out},\mathbf{r}^{\prime}\in\mathbb{R}^d\setminus\overline{\Omega}. 
	\end{cases}
\end{equation}

Equation~(\ref{external formal solution of potential}) shows that the exterior solution depends continuously on the prescribed background potential $\phi^{\mathrm{bg}}(\mathbf{r})$ through its values on the outer boundary. The exterior magnetic response therefore converges to a unique, outer-surface determined limiting distribution. This concludes the proof.

\section{Numerical Demonstration}\label{sec3}
To demonstrate the geometric invariance, we performed a series of two-dimensional finite element simulations using the \textit{MagnetoDynamics2D} module within the open-source software \textbf{Elmer}~\cite{elmer}. For instructional purposes, commercial finite-element software such as \textbf{COMSOL Multiphysics}, when available, provide a more user-friendly interface and may be more accessible to students. Restricting the simulations to two dimensions allows for computational efficiency while capturing the essential physics of the problem. The computational domain consists of various internal structures sharing a common outer boundary, embedded in vacuum. 

For the two-dimensional magnetization problem, we show in the Appendix section that the external magnetic scalar potential admits a multipole expansion associated with the magnetized body:
\begin{equation}
	\begin{aligned}
		\phi^+(\mathbf{r})&=\phi^{bg}(\mathbf{r})+\phi_{\mathrm{mono}}+\phi_{\mathrm{dipole}}+\phi_{\mathrm{quadrupole}}+\cdots\\
		&=\phi^{bg}(\mathbf{r})+\quad\frac{1}{2\pi}\left[0+\frac{\mathbf{r}\cdot\mathbf{m}}{r^2}+\frac{1}{2}\sum_{i,j}\frac{2r_ir_j}{r^4}Q_{ij}+\cdots\right],
	\end{aligned}
\end{equation}
where 
\begin{equation}
	\mathbf{m}=\int_{\Omega}dS^{\prime}\mathbf{M}(\mathbf{r}^{\prime}),m_i=\int_{\Omega}dS^{\prime}M_i(\mathbf{r}^{\prime}),
\end{equation}
and 
\begin{equation}
	Q_{ij}=\int_{\Omega}dS^{\prime}\left[M_i(\mathbf{r}^{\prime})r_j^{\prime}+M_j(\mathbf{r}^{\prime})r_i^{\prime}-\delta_{ij}\mathbf{M}(\mathbf{r}^{\prime})\cdot\mathbf{r}^{\prime}\right].
\end{equation}
Specifically, the symmetric and traceless second-order tensor \(\mathbf{Q}\) in two dimensions has only two degrees of freedom, and can be written as
\begin{equation}
	\mathbf{Q} = \begin{bmatrix} Q_{xx} & Q_{xy} \\ Q_{xy} & -Q_{xx} \end{bmatrix}.
\end{equation}

\begin{figure}[h!]
	\includegraphics[width = 1\textwidth]{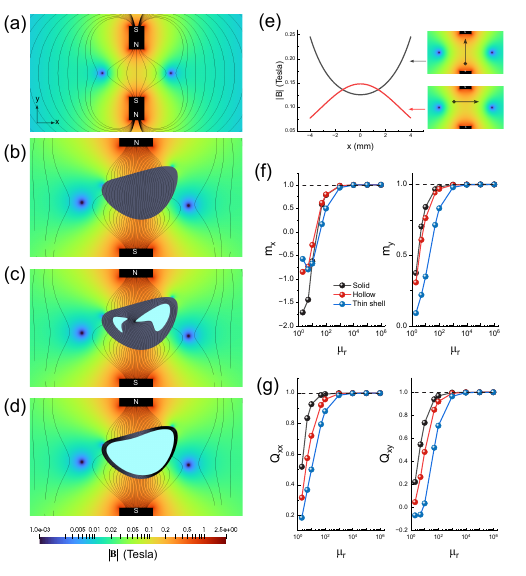}
	\caption{External magnetic response of bodies with identical outer surfaces under a nonuniform magnetic field.
		(a, e) A nonuniform magnetic field generated by a pair of uniformly magnetized bar magnets.
		(b--d) Magnetic flux lines and their density in the exterior of three bodies with identical outer surfaces but differing internal structures, each with a relative permeability of $10^6$ : a solid body (b), a hollow shell (c), and a thin shell (d).
		(f--g) Magnetic dipole vector components (f) and quadrupole tensor components (g) as functions of increasing permeability. All values are normalized to those of the solid body at a relative permeability of $10^6$.}
	\label{Figure03}
\end{figure}

Figures~\ref{Figure02} and~\ref{Figure03} illustrate the magnetization response under uniform and nonuniform external fields, respectively. 
In the uniform-field case (Fig.~\ref{Figure02}), the external flux patterns remain nearly identical across all configurations when the relative permeability is set to $\mu_r = 10^6$, which effectively approximates the infinite-permeability limit. 
The convergence of the magnetic dipole moment and quadrupole tensor with increasing $\mu_r$ is shown in Fig.~\ref{Figure02}(d) and (e), revealing rapid saturation for $\mu_r \gtrsim 10^4$. 
A similar invariance is observed under a spatially varying external field generated by a pair of bar magnets (Fig.~\ref{Figure03}).

\begin{figure}[h!]
	\includegraphics[width = 1\textwidth]{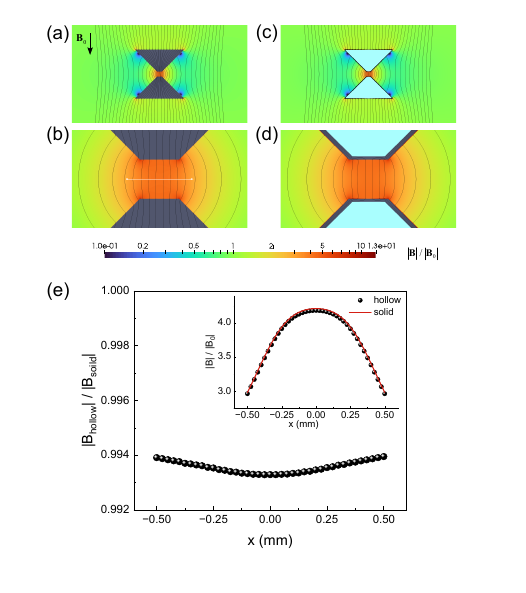}
	\caption{Magnetic flux concentration in solid and hollow MFC structures under a uniform external field.
		(a--d) Magnetic flux lines and their density in the exterior of two magnetic flux concentrators (MFCs), each with a relative permeability of $10^4$: solid MFC with 100\% volume fraction (a,b), and hollow MFC with 8.7\% volume fraction (c,d). 
		(e) Relative magnetic flux density along the central axis of the hollow MFC, normalized to that of the solid MFC. Inset: absolute magnetic flux density along the central axes of both structures.}
	\label{Figure04}
\end{figure}

\section{Application to Lightweight Device Design}\label{Application in MFC}
An immediate application is lightweight device design. A representative example is the magnetic flux concentrator (MFC), which is widely used in magnetic sensors, energy harvesters, and wireless power transfer systems~\cite{10.1063/1.3056152,PhysRevResearch.2.023394,D0EE01574A,10265057}. Conventionally, MFCs are fabricated as fully solid components made of soft magnetic materials with high relative permeability, ranging from $\mu_r \sim 10^{2}$--$10^{4}$ for ferrites to $\mu_r \sim 10^{4}$--$10^{5}$ for mu-metal. The invariance demonstrated here provides a simple yet powerful design principle: when device performance depends on the external magnetic response, a hollow structure can effectively replace its solid counterpart. Such hollowing strategies, however, remain largely unexplored in the existing literature on MFC design, indirectly reflecting the lack of awareness of this invariance.

Figure~\ref{Figure04} illustrates this concept by comparing the magnetic response of a solid MFC (100\% volume fraction) and a thin-shell MFC (8.7\% volume fraction) under a uniform external field. At a high permeability of $\mu_r = 10^4$, the two configurations exhibit virtually identical performance, indicating that significant reductions in material usage and device weight can be achieved without compromising functionality.

As a supplementary remark, we emphasize that the geometric invariance reported here is not applicable to permanent-magnet technologies (e.g., NdFeB magnets, whose cost and availability are strongly influenced by the global rare-earth supply-chain), nor to current-carrying devices such as transformer cores, where device performance is dictated by magnetic flux inside the magnetic material.

\section{Pedagogical Value and Applications}\label{sec5}

The pedagogical value of this invariance in electrodynamics is self-evident. 
In particular, from the perspective of transformations and symmetries, the quasi-equipotential concept describes the geometric invariance of the scalar potential within the interior of a body in the high-permeability limit, while the present invariance characterizes the geometric invariance of the potential in the exterior space. Together, they provide a complete characterization of Laplace transmission problems in the infinite-permeability limit. Accordingly, we believe that this concept holds significant potential for inclusion in future electrodynamics textbooks and courses.

This invariance can be introduced in teaching electromagnetic boundary‐value problems either as an advanced topic--with the complete conclusion and a rigorous proof, or in a simplified form by considering magnetization problems of analytically solvable, highly symmetric bodies (such as a solid sphere and a spherical shell). In the latter case, students can compare the external magnetic fields and total magnetic moments of the two configurations in the high‐permeability limit, which naturally leads to the stated conclusion.

\section{Conclusion}
In summary, we have identified a simple yet robust geometric invariance in transmission problems of the Laplace equation--one governed entirely by a body's outer surface, independent of its internal geometry. This invariance arises not from the irrelevance of the interior, but from the fact that the boundary alone encodes all essential information. In magnetostatics, it complements the prevailing notion that a body's perturbation of the external magnetic field depends on its overall geometry, including internal structures when the permeability is finite. Given the ubiquity of Laplace transmission problems across physics, including heat conduction, electrostatic polarization, and acoustic scattering, this geometric invariance exhibits cross-disciplinary universality. It provides a direct theoretical basis for lightweight device design and can serve as a benchmark for verifying the accuracy of numerical solutions. Most importantly, together with the quasi-equipotential property, it offers a complete characterization of Laplace transmission problems in the infinite permeability limit.

\section*{Author Declarations}
\textbf{Conflict of Interest}

The authors have no conflicts of interest to disclose.

\appendix  
\renewcommand{\theequation}{A\arabic{equation}}
\setcounter{equation}{0}

\section*{Appendix: Mathematical Derivations}\label{Appendix}

\subsection {Proof of the quasi-equipotential and geometric invariance}

In the limit of infinite magnetic permeability within a region \(\Omega\), the normal derivative of the scalar magnetic potential at the boundary satisfies
\begin{equation}
\frac{\partial \phi^-}{\partial \mathbf{n}} = \frac{\mu_0}{\mu} \frac{\partial \phi^+}{\partial \mathbf{n}}\to 0
\end{equation}
implying that the internal problem reduces to an \textit{interior Neumann problem} for the potential \(\phi^-\) with quasi-zero normal derivative on the boundary.
\begin{equation}\label{interior Neumann problem}
	\begin{cases}
		\Delta\phi^-(\mathbf{r})=0,&\mathbf{r}\in\Omega\\
		\frac{\partial\phi^-}{\partial\mathbf{n}}=f(\mathbf{r})\to0,&\mathbf{r}\in\partial\Omega
	\end{cases}
\end{equation}
given the Neumann Green's function which satisfies
\begin{equation}\label{interior Neumann Green's function}
	\begin{cases}
		\Delta G(\mathbf{r}-\mathbf{r}^{\prime})=-\delta(\mathbf{r}-\mathbf{r}^{\prime}),&\mathbf{r},\mathbf{r}^{\prime}\in\Omega\\\frac{\partial G(\mathbf{r}-\mathbf{r}^{\prime})}{\partial\mathbf{n}}=-\frac{1}{\mathrm{A}},&\mathbf{r}\in\partial\Omega,\mathbf{r}^{\prime}\in\Omega,
	\end{cases}
\end{equation}
where A denotes the area of the boundary surface \(\partial\Omega\).  The potential \(\phi^-(\mathbf{r})\) can be expressed in terms of a boundary integral involving the Green's function\cite[Jackson, sec.~1.10]{jackson_classical_1999},
\begin{equation}\label{formal solution of interior potential}
	\begin{aligned}
		\phi^{-}(\mathbf{r})&=\int_{\partial\Omega}dS^{\prime}G(\mathbf{r}-\mathbf{r}^{\prime})\frac{\partial\phi^-}{\partial\mathbf{n}^{\prime}}-\int_{\partial\Omega}dS^{\prime}\phi^-\frac{\partial G(\mathbf{r}-\mathbf{r}^{\prime})}{\partial\mathbf{n}^{\prime}}\\&=\int_{\partial\Omega}dS^{\prime}G(\mathbf{r}-\mathbf{r}^{\prime})f(\mathbf{r}^{\prime})|_{f(\mathbf{r}^{\prime})\to0}+\frac{1}{\mathrm{A}}\int_{\partial\Omega}dS^{\prime}\phi^{-}(\mathbf{r}^{\prime})\\&\to\phi_{0},
	\end{aligned}
\end{equation}
where 
\[\phi_0=\frac{1}{\mathrm{A}}\int_{\partial\Omega}dS^{\prime}\phi^-(\mathbf{r}^{\prime}).\]

Equation~(\ref{formal solution of interior potential}) show that the internal potential tends to a constant.  Since Neumann problems admit solutions only up to an additive constant, $\phi_0 + C$ also satisfies the same boundary formulation. However, in the presence of a background field $\phi^{\mathrm{bg}}(\mathbf{r})$, shifting $\phi_0$ by $C$ implies a corresponding shift in $\phi^{\mathrm{bg}}$. Without loss of generality, we set $\phi_0 = 0$, so the interior potential approaches zero. By continuity, the surface potential on $\partial\Omega$ also tends to zero, reducing the exterior problem to an \textit{exterior Dirichlet problem}  with quasi-zero boundary values on \(\partial\Omega\).

\begin{equation}\label{exterior Dirichlet problem_Appendixes}
	\begin{cases}
		\Delta\phi^+(\mathbf{r})=0,&\mathbf{r}\in\mathbb{R}^d\setminus\overline{\Omega}\\ \phi^+(\mathbf{r})\to0,&\mathbf{r}\in\partial\Omega^{out}\\\phi^+(\mathbf{r})\to\phi^{bg}(\mathbf{r}),&\left|\mathbf{r}\right|\to\infty
	\end{cases}
\end{equation}
Let \(\phi^+(\mathbf{r})=u(\mathbf{r})+\phi^{bg}(\mathbf{r})\), the perturbation term \(u(\mathbf{r})\) satisfies
\begin{equation}\label{perturbation term}
	\begin{cases}\Delta u(\mathbf{r})=0,&\mathbf{r}\in\mathbb{R}^d\setminus\overline{\Omega}
		\\u(\mathbf{r})\to-\phi^{bg},&\mathbf{r}\in\partial\Omega^{out}
		\\u(\mathbf{r})\to0,&|\mathbf{r}|\to\infty.
	\end{cases}
\end{equation}
Given the Dirichlet Green's function which satisfies
\begin{equation}
	\begin{cases}
		\Delta G(\mathbf{r}-\mathbf{r}^{\prime})=-\delta(\mathbf{r}-\mathbf{r}^{\prime}),\quad\mathbf{r},\mathbf{r}^{\prime}\in\mathbb{R}^d\setminus\overline{\Omega}\\
		G(\mathbf{r}-\mathbf{r}^{\prime})=0,\quad\mathbf{r}\in\partial\Omega^{out},\mathbf{r}^{\prime}\in\mathbb{R}^d\setminus\overline{\Omega}\\
		G(\mathbf{r}-\mathbf{r}^{\prime})\to0,\quad|\mathbf{r}-\mathbf{r}^{\prime}|\to\infty,\mathbf{r}\in\partial\Omega^{out},\mathbf{r}^{\prime}\in\mathbb{R}^d\setminus\overline{\Omega}. 
	\end{cases}
\end{equation}
the formal solution of the equations \ref{perturbation term} can be composed of the integral of Green's function on the boundary,

\begin{equation}\label{formal solution of exterior potentia_Appendixes}
	\begin{aligned}
		u(\mathbf{r})&=\int_{\partial\Omega^{out}}dS^{\prime}G(\mathbf{r}-\mathbf{r}^{\prime})\frac{\partial\phi^{+}}{\partial\mathbf{n}^{\prime}}-\int_{\partial\Omega^{out}}dS^{\prime}u(\mathbf{r}^{\prime})\frac{\partial G(\mathbf{r}-\mathbf{r}^{\prime})}{\partial\mathbf{n}^{\prime}}
		\\&=-\int_{\partial\Omega^{out}}dS^{\prime}u(\mathbf{r}^{\prime})\frac{\partial G(\mathbf{r}-\mathbf{r}^{\prime})}{\partial\mathbf{n}^{\prime}}
		\\&\to\int_{\partial\Omega^{out}}dS^{\prime}\phi^{bg}(\mathbf{r}^{\prime})\frac{\partial G(\mathbf{r}-\mathbf{r}^{\prime})}{\partial\mathbf{n}^{\prime}}.\end{aligned}
\end{equation}
Equation~(\ref{formal solution of exterior potentia_Appendixes}) shows that the exterior solution depends continuously on the prescribed background potential $\phi^{\mathrm{bg}}(\mathbf{r})$ through its values on the outer boundary. Hence, the exterior potential \(\phi^+\) converges to a unique limiting configuration, thereby establishing the geometric invariance discussed in this paper. 

\subsection{Integral expressions for the dipole and quadrupole}
On the other hand, the perturbation term  \(u(\mathbf{r})\) can be expanded in terms of magnetic multipole moments associated with the magnetized body:
\begin{equation}
	u(\mathbf{r})=\phi_{\mathrm{mono}}+\phi_{\mathrm{dipole}}+\phi_{\mathrm{quadrupole}}+\cdots~.
\end{equation}

The perturbation term \( u(\mathbf{r}) \) is entirely determined by the magnetization distribution within the magnetized body. From the perspective of equivalent magnetic charges, and by direct analogy with the electrostatic problem, \( u(\mathbf{r}) \) arises from the combined contributions of volume magnetic charges and surface magnetic charges on the boundary~\cite[Jackson, page 197, equation 5.100]{jackson_classical_1999}.
\subsubsection{3D case}
In the three-dimensional case,
\begin{equation}
	u(\mathbf{r})=\frac{1}{4\pi}\int_{\Omega}dV^{\prime}\frac{\rho_m}{\left|\mathbf{r}-\mathbf{r}^{\prime}\right|}+\frac{1}{4\pi}\int_{\partial\Omega}dS^{\prime}\frac{\sigma_m}{\left|\mathbf{r}-\mathbf{r}^{\prime}\right|}.
\end{equation}
By substituting \(\rho_m=-\nabla \cdot \mathbf{M}(\mathbf{r}) \) and \(\sigma_m=\mathbf{n}\cdot\mathbf{M}\),

\begin{equation}
	\begin{aligned}
		u(\mathbf{r})&=-\frac{1}{4\pi}\int_{\Omega}dV^{\prime}\frac{\nabla^{\prime}\cdot\mathbf{M}(\mathbf{r}^{\prime})}{|\mathbf{r}-\mathbf{r}^{\prime}|}+\frac{1}{4\pi}\int_{\partial\Omega}dS^{\prime}\frac{\mathbf{n}^{\prime}\cdot\mathbf{M}(\mathbf{r}^{\prime})}{|\mathbf{r}-\mathbf{r}^{\prime}|}\\&=\frac{1}{4\pi}\int_{\Omega}dV^{\prime}\mathbf{M}(\mathbf{r}^{\prime})\cdot\nabla^{\prime}\frac{1}{|\mathbf{r}-\mathbf{r}^{\prime}|}.
	\end{aligned}
\end{equation}
Using the expansion
\begin{equation}
	\frac{1}{\left|\mathbf{r}-\mathbf{r}^{\prime}\right|}=\frac{1}{r}-\mathbf{r}^{\prime}\cdot\nabla\frac{1}{r}+\frac{1}{2}\left(\mathbf{r}^{\prime}\cdot\nabla\right)^2\frac{1}{r}+\cdots~,
\end{equation}
we can derive the expressions for the multipole terms in \( u(\mathbf{r}) \) as follows. Here and below, we adopt the Einstein summation convention: repeated indices are implicitly summed over the corresponding range.
\begin{equation}
	\phi_{\mathrm{mono}}=\frac{1}{4\pi}\int_{\Omega}dV^{\prime}\mathbf{M}(\mathbf{r}^{\prime})\cdot\nabla^{\prime}\frac{1}{r}=0,
\end{equation}

\begin{equation}
	\begin{aligned}
		\phi_{\mathrm{dipole}}&=\frac{1}{4\pi}\int_\Omega dV^{\prime}\mathbf{M}(\mathbf{r}^{\prime})\cdot\nabla^{\prime}\left[-\mathbf{r}^{\prime}\cdot\nabla\frac{1}{r}\right]\\&=\frac{1}{4\pi}\int_{\Omega}dV^{\prime}M_{i}(\mathbf{r}^{\prime})\frac{\partial}{\partial r_{i}^{\prime}}\left[-r_{\alpha}^{\prime}\frac{\partial}{\partial r_{\alpha}}\frac{1}{r}\right]\\&=\frac{1}{4\pi}\int_\Omega dV^{\prime}M_i(\mathbf{r}^{\prime})\frac{r_i}{r^3}\\&=\frac{1}{4\pi}\frac{r_i}{r^3}\int_\Omega dV^{\prime}M_i(\mathbf{r}^{\prime})\\&=\frac{1}{4\pi}\frac{r_i}{r^3}m_i\\&=\frac{1}{4\pi}\frac{\mathbf{r}\cdot\mathbf{m}}{r^3},
	\end{aligned}
\end{equation}

where dipole moment 
\begin{equation}
	\mathbf{m}=\int_{\Omega}dV^{\prime}\mathbf{M}(\mathbf{r}^{\prime}),\quad m_i=\int_{\Omega}dV^{\prime}M_i(\mathbf{r}^{\prime}).
\end{equation}

\begin{equation}
	\begin{aligned}
		\phi_{\mathrm{quadrupole}}&=\frac{1}{4\pi}\int_{\Omega}dV^{\prime}\mathbf{M}(\mathbf{r}^{\prime})\cdot\nabla^{\prime}\left[\frac{1}{2}\left(\mathbf{r}^{\prime}\cdot\nabla\right)^{2}\frac{1}{r}\right]\\&=\frac{1}{4\pi}\int_\Omega dV^{\prime}M_i(\mathbf{r}^{\prime})\frac{\partial}{\partial r_i^{\prime}}\left[\frac{1}{2}r_\alpha^{\prime}r_\beta^{\prime}\frac{\partial^2}{\partial r_\alpha\partial r_\beta}\frac{1}{r}\right]\\&=\frac{1}{4\pi}\int_{\Omega}dV^{\prime}M_{i}(\mathbf{r}^{\prime})\frac{\partial}{\partial r_{i}^{\prime}}\left[\frac{1}{2}r_{\alpha}^{\prime}r_{\beta}^{\prime}\left(\frac{3r_{\alpha}r_{\beta}-\delta_{\alpha\beta}r^{2}}{r^{5}}\right)\right]\\&=\frac{1}{4\pi}\cdot\frac{1}{2}\left(\frac{3r_\alpha r_\beta-\delta_{\alpha\beta}r^2}{r^5}\right)\int_\Omega dV^{\prime}M_i(\mathbf{r}^{\prime})\frac{\partial}{\partial r_i^{\prime}}\left(r_\alpha^{\prime}r_\beta^{\prime}\right)\\&=\frac{1}{4\pi}\cdot\frac{1}{2}\sum_{\alpha,\beta}\left(\frac{3r_{\alpha}r_{\beta}-\delta_{\alpha\beta}r^{2}}{r^{5}}\right)\overline{Q}_{\alpha\beta},
	\end{aligned}
\end{equation}
where 
\[
\begin{aligned}
	\bar{Q}_{\alpha\beta}&=\int_{\Omega}dV^{\prime}M_{i}(\mathbf{r}^{\prime})\frac{\partial}{\partial r_{i}^{\prime}}\left(r_{\alpha}^{\prime}r_{\beta}^{\prime}\right)\\&=\int_{\Omega}dV^{\prime}\left[M_{\alpha}(\mathbf{r}^{\prime})r_{\beta}^{\prime}+M_{\beta}(\mathbf{r}^{\prime})r_{\alpha}^{\prime}\right].
\end{aligned}
\]
Define 
\[
\mathrm{Tr}(\overline{Q})=2\int_{\Omega}dV^{\prime}\left[\mathbf{M}(\mathbf{r}^{\prime})\cdot\mathbf{r}^{\prime}\right].
\]
Removing the trace from \(\bar{Q}\) to construct the traceless tensor \(Q\):
\begin{equation}
	\begin{aligned}Q_{ij}&=\overline{Q}_{ij}-\frac{1}{3}\delta_{ij}\mathrm{Tr}(\overline{Q})\\&=\int_{\Omega}dV^{\prime}\left[M_{i}(\mathbf{r}^{\prime})r_{j}^{\prime}+M_{j}(\mathbf{r}^{\prime})r_{i}^{\prime}-\frac{2}{3}\delta_{ij}\mathbf{M}(\mathbf{r}^{\prime})\cdot\mathbf{r}^{\prime}\right].
	\end{aligned}
\end{equation}
Then
\begin{equation}
	\begin{aligned}
		\phi_{\mathrm{quadrupole}}&=\frac{1}{4\pi}\cdot\frac{1}{2}\sum_{i,j}\left(\frac{3r_ir_j-\delta_{ij}r^2}{r^5}\right)\left(Q_{ij}+\frac{1}{3}\delta_{ij}\operatorname{Tr}(\overline{Q})\right)\\&=\frac{1}{4\pi}\cdot\frac{1}{2}\sum_{i,j}\frac{3r_ir_j}{r^5}Q_{ij}.
	\end{aligned}
\end{equation}
From the above, we obtain
\begin{equation}
	\begin{aligned}
		u(\mathbf{r})&=\phi_{\mathrm{mono}}+\phi_{\mathrm{dipole}}+\phi_{\mathrm{quadrupole}}+\cdots\\
		&=\quad\frac{1}{4\pi}\left[0+\frac{\mathbf{r}\cdot\mathbf{m}}{r^3}+\frac{1}{2}\sum_{i,j}\frac{3r_ir_j}{r^5}Q_{ij}+\cdots\right].
	\end{aligned}
\end{equation}
Higher-order multipole terms involve increasingly cumbersome expressions and are not presented here for brevity.
\subsubsection{2D case}
In the two-dimensional case,
\begin{equation}
	\begin{aligned}
		u(\mathbf{r})&=-\frac{1}{2\pi}\int_{\Omega}dS^{\prime}\left[-\nabla^{\prime}\cdot\mathbf{M}(\mathbf{r}^{\prime})\right]\ln\left|\mathbf{r}-\mathbf{r}^{\prime}\right|\\
		&\quad-\frac{1}{2\pi}\int_{\partial\Omega}dl^{\prime}\mathbf{n}^{\prime}\cdot\mathbf{M}(\mathbf{r}^{\prime})\ln\left|\mathbf{r}-\mathbf{r}^{\prime}\right|\\
		&=-\frac{1}{2\pi}\int_{\Omega}dS^{\prime}\mathbf{M}(\mathbf{r}^{\prime})\cdot\nabla^{\prime}\left[\ln|\mathbf{r}-\mathbf{r}^{\prime}|\right].
	\end{aligned}
\end{equation}
Proceeding similarly, we derive
\begin{equation}
	\begin{aligned}u(\mathbf{r})&=\phi_{\mathrm{mono}}+\phi_{\mathrm{dipole}}+\phi_{\mathrm{quadrupole}}+\cdots\\&=\quad\frac{1}{2\pi}\left[0+\frac{\mathbf{r}\cdot\mathbf{m}}{r^2}+\frac{1}{2}\sum_{i,j}\frac{2r_ir_j}{r^4}Q_{ij}+\cdots\right],
	\end{aligned}
\end{equation}
where 
\[
\mathbf{m}=\int_{\Omega}dS^{\prime}\mathbf{M}(\mathbf{r}^{\prime}),m_i=\int_{\Omega}dS^{\prime}M_i(\mathbf{r}^{\prime}),
\]
and 
\begin{equation}
	Q_{ij}=\int_{\Omega}dS^{\prime}\left[M_i(\mathbf{r}^{\prime})r_j^{\prime}+M_j(\mathbf{r}^{\prime})r_i^{\prime}-\delta_{ij}\mathbf{M}(\mathbf{r}^{\prime})\cdot\mathbf{r}^{\prime}\right].
\end{equation}
Specifically, a symmetric and traceless second-order tensor \(\mathbf{Q}\) in two dimensions has only two degrees of freedom, and can be written as
\[
\mathbf{Q} = \begin{bmatrix} Q_{xx} & Q_{xy} \\ Q_{xy} & -Q_{xx} \end{bmatrix}.
\]

\FloatBarrier

\bibliographystyle{unsrt}
\bibliography{reference.bib}

\end{document}